\RequirePackage{lineno}
\documentclass[twocolumn,aps,prc,showpacs,superscriptaddress,preprintnumbers,floatfix,nofootinbib]{revtex4}
\usepackage{epsfig,graphics}
\usepackage{graphicx}% Include figure files
\usepackage{dcolumn}% Align table columns on decimal point
\usepackage{bm}% bold math
\usepackage{amsmath}
\usepackage[usenames]{color}
\usepackage{ulem} %% for strike-through

\begin{document}

\title{Decomposition of jet fragmentation function in high-energy heavy-ion collisions}

\author{Guo-Liang Ma}
\affiliation{Shanghai Institute of Applied Physics, Chinese
Academy of Sciences, P.O. Box 800-204, Shanghai 201800, China}

%\date{\today}

\begin{abstract}
Based on a multi-phase transport model, the measured jet fragmentation function ratio of Pb+Pb collisions to p+p collisions in CERN Large Hadron Collider experiments is decomposed into two parts, corresponding to the two contributions of jet hadronization from fragmentation and coalescence. The results suggest an existence of distinct competitions between two jet hadronization mechanisms for different $\xi$=ln(1/$z$) ranges in different centrality bins. The jet fragmentation functions for different types of hadrons (mesons and baryons) are proposed as a good probe to study the competition between fragmentation and coalescence for the jet hardonization in high-energy heavy-ion collisions.
\end{abstract}


\maketitle

A jet, produced by initial QCD hard scatterings, is one of important probes for studying the properties of strongly interacting matter because it interacts with the medium and loses energy during its traverse~\cite{Whitepapers1,Whitepapers2}. The energy loss results in a medium modification of jet fragmentation function with respect to that in vacuum, which leads to a phenomenon called jet quenching~\cite{Wang:1991xy}.  The phenomenon can be verified by measuring the modifications of jet yields and jet properties.  The measurement of jet fragmentation function can provide a relatively direct way to compare data with theoretical models of jet quenching. The CERN Large Hadron Collider (LHC) measurements of the modification ratio of the jet fragmentation function in Pb+Pb collisions to that in p+p collisions basically show the interesting features of no modification at low $\xi$=ln(1/$z$), suppression at intermediate $\xi$, and enhancement at high $\xi$ for associated charged hadrons inside the jet cone~\cite{Chatrchyan:2012gw,CMS:2012wxa, ATLAS:2012ina}. The measurements cover a wide $\xi$ range, which corresponds to a large $p_{T}$ range from $\sim$ 1 GeV/$c$  to very high $p_{T}$.  Some of theoretical attempts have been made to interpret the modifications. Zapp $et~al.$~\cite{Zapp:2012ak} can basically describe jet fragmentation functions and many other jet-related observables at LHC based on a development of a novel description of jet quenching and its implementation into the Monte Carlo generator JEWEL with the parameters fixed by BNL Relativistic Heavy Ion Collider (RHIC) data. Wang and Zhu~\cite{Wang:2013cia} found a big enhancement at small $z$ for a $\gamma$-tagged jet owing mostly to contributions from radiated gluons within a linearized Boltzmann transport model for jet propagation that includes both elastic parton scattering and induced gluon emission. Kharzeev and Loshaj argue that the suppression of the in-medium fragmentation at intermediate $\xi$ is attributable to the partial screening of the color charge of the jet by the co-moving medium-induced gluon within an effective 1+1 -dimensional quasi-Abelian model~\cite{Kharzeev:2012re}. However, the jet hadronization in these models is based on either the assumption of parton-hadron duality or Lund string fragmentation. Actually, there already exists some experimental evidence for different hadronization mechanisms dominated for different $p_{T}$ ranges in high-energy heavy-ion collisions. For instance, the power-law spectra at high $p_T$ indicates that hadron production is dominated by fragmentation scheme for high transverse momentum ($p_{T}$) range~\cite{Adler:2006bw}. The mechanism of coalescence is preferred for the intermediate-$p_{T}$ range by many experimental observations, such as number of constituent quarks scaling in elliptic flow~\cite{Adams:2005zg,Adare:2006ti} and large ratio of protons over pions~\cite{Abelev:2006jr}.  Hwa and Yang proposed that the jet fragmentation process can be expressed as the recombination of jet shower partons  and the medium partons~\cite{Hwa:2003ic,Hwa:2004ng}. Owing to the recombination of thermal and shower partons, the structure of jets produced in high-energy heavy-ion collisions should be different from that in p+p collisions. In this paper, the effect of both hadronization mechanisms, including fragmentation and coalescence, on the medium modification of jet fragmentation function is investigated based on a multi-phase transport (AMPT) model. The measured jet fragmentation function is found to have two contribution components from fragmentation and coalescence, which can bring a medium-induced enhancement of baryon yield inside a jet.

The AMPT model with the string-melting scenario, which has well described many experimental observables~\cite{Lin:2004en,Chen:2006ub, Zhang:2005ni, Ma:2011uma, Ma:2010dv}, is implemented in this work. It consists of four main stages of high-energy heavy-ion collisions: the initial condition, parton cascade, hadronization, and hadronic rescatterings. The initial condition, which includes the spatial and momentum distributions of minijet partons and soft string excitations, is obtained from the HIJING model~\cite{Wang:1991hta,Gyulassy:1994ew}. In the string-melting scenario of AMPT model, the minijet partons and soft strings are fragmented into hadrons with the LUND fragmentation, built in the PYTHIA routine~\cite{Sjostrand:1993yb}, and then these hadrons are converted into on-shell quarks and anti-quarks according to the flavor and spin structures of their valence quarks. The evolution of partonic plasma, parton cascade, is simulated by Zhang's parton cascade (ZPC) model~\cite{Zhang:1997ej}, in which the partonic cross section is controlled by the value of a strong coupling constant and the Debye screening mass. The process of parton cascade only includes two-body elastic collisions at present. For hadronization, partons are converted into hadrons with a simple coalescence model by combining the nearest partons into mesons and baryons. It conserves the three-momentum during coalescence and determines the hadron species according to the flavor and invariant mass of coalescing partons. Then the dynamics of the subsequent hadronic rescatterings is described by a relativistic transport (ART) model~\cite{Li:1995pra}.  In this work, the AMPT model with the newly fitted parameters, as listed in Ref.~\cite{Xu:2011fi}, is used to simulate Pb+Pb collisions at $\sqrt{s_{_{\rm NN}}}$ = 2.76 TeV, which has shown good descriptions for many experimental observables at LHC energy, such as pseudorapidity and $p_{T}$ distributions~\cite{Xu:2011fi} and harmonic flows~\cite{Xu:2011fe, Xu:2011jm}.  Two sets of partonic interaction cross sections, 0 and 1.5 mb, are applied to simulate two different physical scenarios for hadronic interactions only and parton + hadronic interactions, respectively.

To study the jet energy loss behaviors, a dijet of $p_{T}\sim$ 90 GeV/$c$ is triggered with the jet-triggering technique in the HIJING model,  because the production cross section of dijet is quite small especially for large transverse momentum. Several hard dijet production processes with high virtualities are additionally taken into account in the initial condition of the AMPT model, including $q_{1}+q_{2}\rightarrow q_{1}+q_{2}$, $q_{1}+\bar{q_{1}}\rightarrow q_{2}+\bar{q_{2}}$, $q+\bar{q}\rightarrow g+g$, $q+g\rightarrow q+g$, $g+g\rightarrow q+\bar{q}$, and $g+g\rightarrow g+g$~\cite{Sjostrand:1993yb}. The high-$p_{T}$ primary partons pullulate to jet showers full of lower virtuality partons through initial- and final- state QCD radiations. The jet parton showers are converted into clusters of on-shell quarks and anti-quarks through the string-melting mechanism of AMPT model. In a sense, the melting scenario for jets, which bears some analogy to the medium-induced subsequent radiations, happens before jet-medium interactions in the logical structure of the AMPT model. After the melting process, not only a quark and anti-quark plasma is formed, but also jet quark showers are built up, therefore the initial configuration between dijet and the medium is ready to interact. In the following, the ZPC model automatically simulates all possible elastic partonic interactions among medium partons, between jet shower partons and medium partons, and among jet shower partons, but without including inelastic parton interactions or further radiations at present. Two sets of partonic interaction cross sections, 0 or 1.5 mb, can be used to turn off or on the process of parton cascade to see the effect of jet-medium interactions in this study. When the partons freeze out, they are recombined into medium hadrons or jet shower hadrons via the simple coalescence model. The formed jet shower hadrons include the recombinations among shower partons and between shower partons and medium partons. The final-state hadronic interactions between jet shower hadrons and hadronic medium can be described by the ART model. Recently, the model has successfully given some qualitative descriptions to the experimental results from full jet reconstruction at LHC, such as $\gamma$-jet imbalance~\cite{Ma:2013bia} and dijet asymmetry~\cite{Ma:2013pha}.

To acquire jet fragmentation function, the kinetic cuts for jet reconstruction are chosen as same as in the CMS experiment~\cite{CMS:2012wxa}. The jet cone size is set to be 0.3. The transverse momentum of jet is required to be larger than 100 GeV/$c$ ($p_{T} >$ 100 GeV/$c$) within a pseudorapidity range of $0. 3 < |\eta|<2$ for this analysis, where jets within $|\eta|<0.3$ are excluded to avoid the overlap between the signal jet region and the jet background estimation region. An anti-$k_{t}$ algorithm from the standard Fastjet package is used to reconstruct full jets~\cite{Cacciari:2011ma}. Jet fragmentation function is obtained by correlating charged hadrons with $p_{T}>$ 1 GeV/$c$ falling within the jet cone, with respect to the axis of reconstructed jet. As the CMS experiment defined, the jet fragmentation function, D($\xi$)=$1/N_{jet}dN_{ch}/d\xi$, can be presented as a function of the variable $\xi = ln(1/z)$, where $z = p_{||}^{ch}/p^{jet}$ is the fraction of the jet energy carried by the charged particle, $p_{||}^{ch}$ is the momentum component of charged particle along the jet axis, $p^{jet}$ is the magnitude of reconstructed jet momentum, and $N_{jet}$ is total number of jets. All charged particles in the cone of 0.3 around jet axis are included in this analysis. It should be noted that lower $\xi$ actually corresponds to higher $p_{T}$.  An $\eta$-reflection method as used in the CMS experiment, i.e. selecting charged particles that lie in a background jet cone obtained by reflecting the original jet cone around $\eta$=0 while keeping the same $\phi$ coordinate, is used to estimate the background, which is subtracted from the reconstructed jet fragmentation function in Pb+Pb collisions. 

\begin{figure}
\includegraphics[scale=0.45]{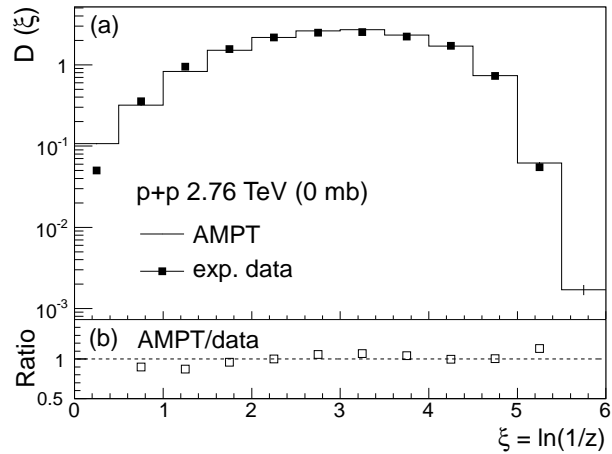}
\caption{(a) The jet fragmentation function D($\xi$) in p+p 2.76-TeV, where the histograms represent the AMPT result with hadronic interactions only (0 mb), and the squares represent the data from the CMS experiment~\cite{CMS:2012wxa}. (b) The ratios of AMPT result to experimental data.}
 \label{fig-pp}
\end{figure}

Figure~\ref{fig-pp} (a) shows the jet fragmentation function D($\xi$) in p+p 2.76-TeV. Form a quantitative comparison of the ratios of AMPT result to experimental data shown in Figure~\ref{fig-pp} (b), the results obtained from AMPT simulations with hadronic interactions only basically can describe the jet fragmentation function in p+p collisions, which provides a reliable baseline for the following studies about those in Pb+Pb collisions with the AMPT model with both partonic and hadronic interactions.

\begin{figure}
\includegraphics[scale=0.45]{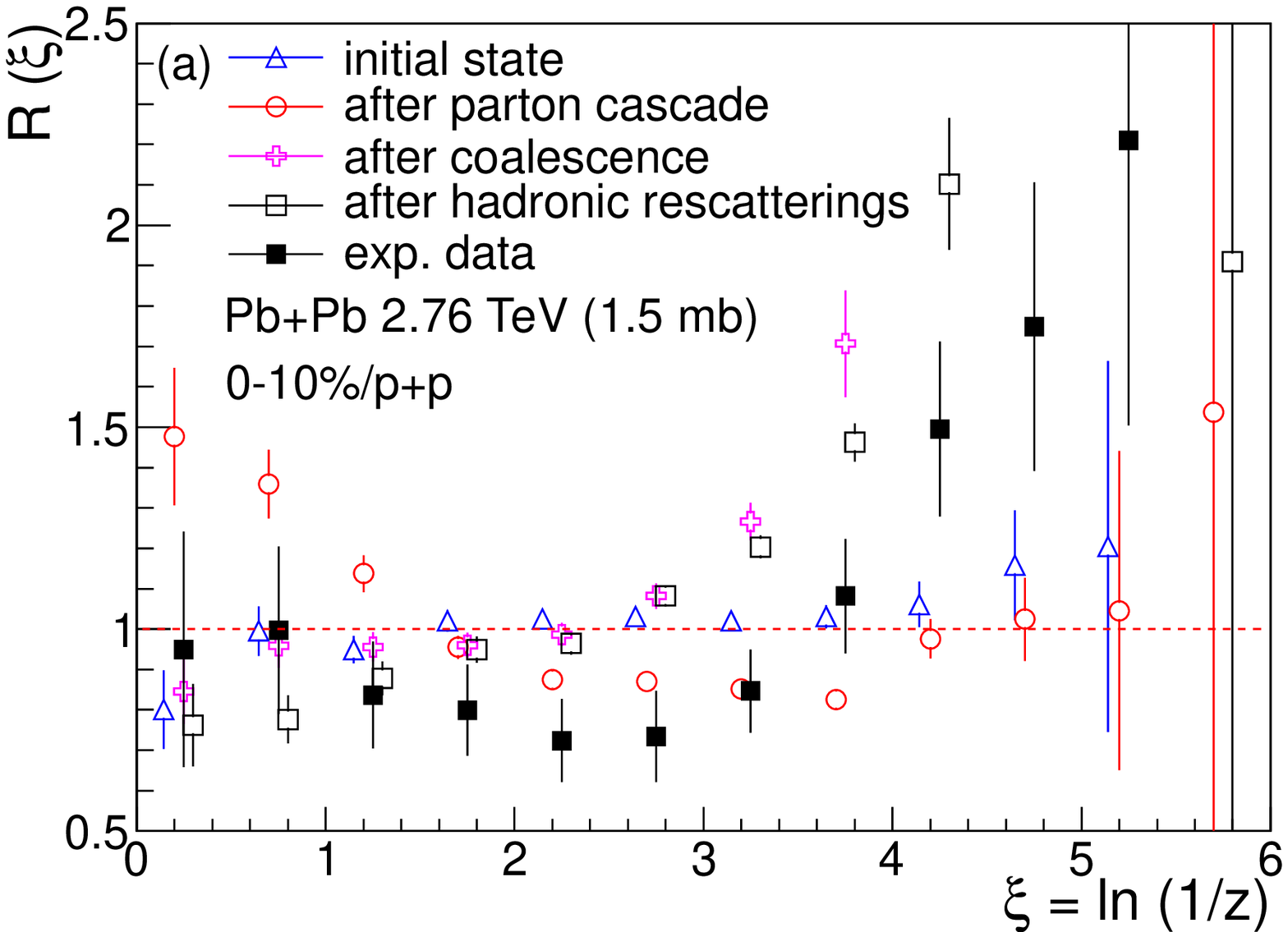}
\includegraphics[scale=0.45]{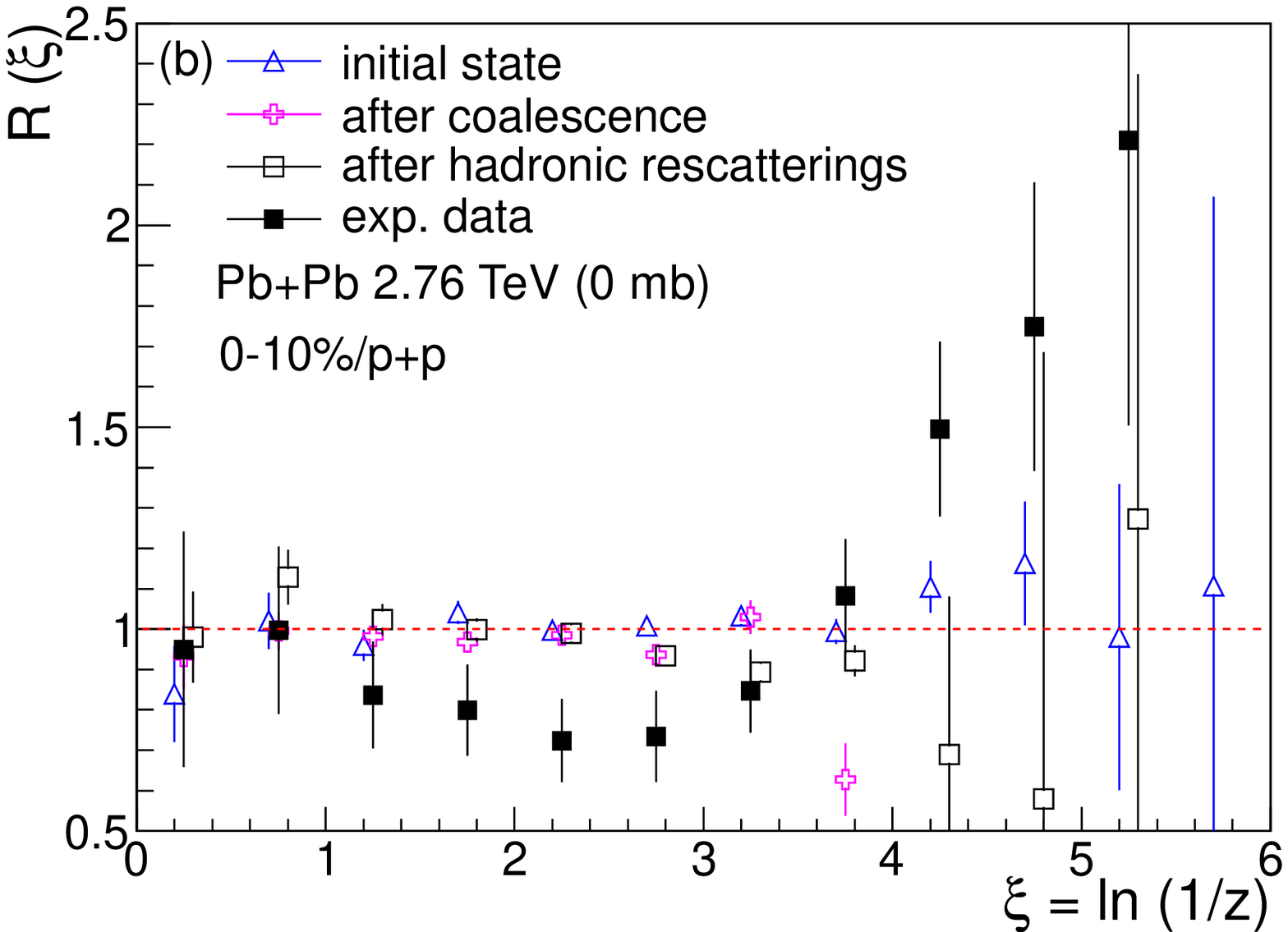}
\caption{(Color online) The jet fragmentation function ratios of the most central centrality bin (0-10\%) in Pb+Pb 2.76-TeV collisions to p+p collisions at different evolution stages, where plots (a) and (b) represent  the AMPT results with partonic+hadronic and hadronic interactions only respectively. The solid squares represent the data from the CMS experiment~\cite{CMS:2012wxa}. Some points are slightly shifted along the $x$ axis for better representation.
}
 \label{fig-fragevo}
\end{figure}

Because heavy-ion collisions actually are dynamical evolutions including many important stages, the evolution course of jet fragmentation function can provide important information about the mechanism of medium modification of jet fragmentation function in Pb+Pb collisions.   Figure~\ref{fig-fragevo} (a) and (b) present the jet fragmentation function ratios of the most central centrality bin (0-10\%) in Pb+Pb collisions to p+p collisions, i.e. R($\xi$)=$D_{Pb+Pb}(\xi)/ D_{p+p}(\xi)$, at different evolution stages from AMPT simulations with partonic+hadronic (1.5 mb) and hadronic interactions only (0 mb), respectively. In Figure~\ref{fig-fragevo} (a), the initial jet fragmentation function ratio is around unity which indicates no modification in the initial state of Pb+Pb collisions. Two basic features of modification, an enhancement at low $\xi$ and a suppression at intermediate $\xi$, appear in jet fragmentation function ratio after parton cascade process. The enhancement is because the energy loss of jet is more significant than that of leading-like partons, which relatively decrease their $\xi$. However, the suppression is the result of the decrease of associated particles with intermediate $p_{T}$ owing to the jet energy loss in the partonic medium, which possibly are shifted to lower $p_T$ or even thermalized. However, the expected high-$\xi$ enhancement owing to the shift or thermalization is hard to be seen for the current statistics. A significant enhancement around intermediate and high $\xi$ and small suppression at low $\xi$ are observed after coalescence. It is because the coalescence mechanism in the AMPT model increases the total momentum of jet a little, owing to the involution of medium partons, and also increases the momenta of shower hadrons in comparison with the previous stage. The final-state hadronic rescatterings do not seem to change the formed jet fragmentation function ratio any more. In Figure~\ref{fig-fragevo} (b), jet fragmentation function ratios from different stages of Pb+Pb collisions in the AMPT model with hadronic interactions only are always consistent with unity, which indicates no obvious modification with respect to p+p collisions if without partonic interactions in Pb+Pb collisions. However, neither of the final jet fragmentation function ratios in the two sets of simulations can fit the experimental data for the whole $\xi$ range. 

As introduced above, the coalescence mechanism is thought to be a dominant way of hadronization for intermediate $\xi$ (i.e. intermediate $p_{T}$), whereas the fragmentation mechanism takes over for low $\xi$ (i.e. high $p_{T}$).  Actually, the interplay of fragmentation and coalescence indeed can give very good descriptions about $p_{T}$ spectra and elliptic flow in a wide $p_{T}$ range~\cite{Fries:2003kq}. The reason the AMPT results can not match the measured jet fragmentation function ratio for the whole $\xi$ range is that the AMPT model with string-melting scenario only uses a coalescence model for hadronization, but misses the other important one, i.e. fragmentation. To well describe the experimental data of jet fragmentation function ratio in the whole $\xi$ range, it is proposed to decompose the measured jet fragmentation function ratio to,
\begin{equation}
R(\xi)=\lambda_{f}R_{f}(\xi)+\lambda_{c}R_{c}(\xi),
\label{eq1}
\end{equation}
where $\lambda_{f}R_{f}(\xi)$ and $\lambda_{c}R_{c}(\xi)$ are fragmentation part and coalescence part respectively, which are assumed to coexist in the measured jet fragmentation function ratio $R(\xi)$, $\lambda_{f}$ and $\lambda_{c}$ are the contribution factors for fragmentation part and coalescence part respectively. The functional form of  $R_{f}(\xi)$ is assumed to be as same as that of jet fragmentation function ratio after parton cascade, based on the parton-hadron duality or the subleading correction effect of fragmentation on the nuclear modification factor~\cite{Renk:2007zz}. The functional form of  $R_{c}(\xi)$ is assumed to be as same as from jet fragmentation function ratio after hadronic rescatterings, which includes both effects of coalescence and hadronic rescatterings. Thus, the two contribution parts can be obtained by fitting the experimental data of $R(\xi)$ with Equation~(\ref{eq1}). It should be noted that $\lambda_{f}$ and $\lambda_{c}$ are assumed to be independent of $\xi$ for simplicity in this work, which also can be understood as the averaged values.

\begin{figure}
\includegraphics[scale=0.45]{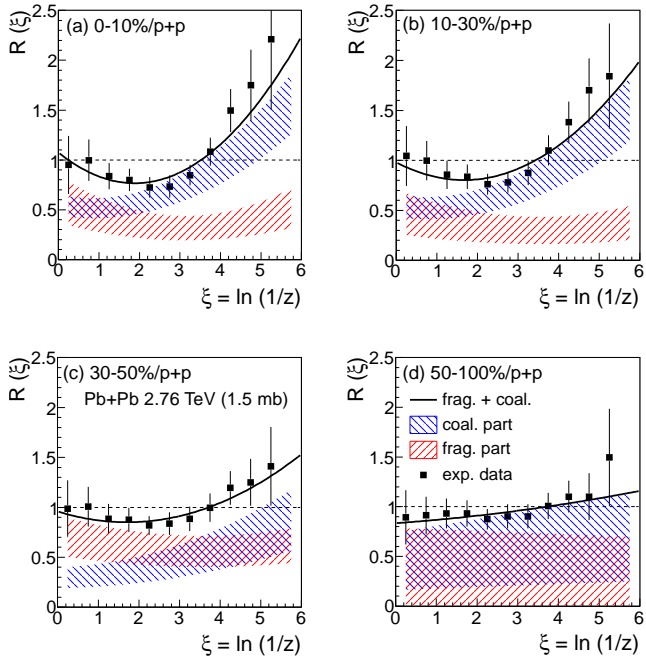}
\caption{(Color online) The jet fragmentation function ratios of different centrality bins in Pb+Pb 2.76-TeV collisions to p+p collisions. The solid curves show two-component (fragmentation + coalescence) fitting functions, while the two kinds of hatched areas give the fragmentation and coalescence contribution parts, i.e. $\lambda_{f}R_{f}(\xi)$ and $\lambda_{c}R_{c}(\xi)$, for the jet fragmentation function ratios measured by the CMS experiment~\cite{CMS:2012wxa}. 
}
\label{fig-fit}
\end{figure}

The solid curves in Figures~\ref{fig-fit} (a)-(d) show the combinational fittings to the measured jet fragmentation function ratios of different centrality bins in Pb+Pb collisions to p+p collisions with Equation~(\ref{eq1}), and Table~\ref{tab:para} lists the fitting parameters of $\lambda_{f}$ and $\lambda_{c}$.  From the fittings, the two contributions parts from fragmentation and coalescence, $\lambda_{f}R_{f}(\xi)$ and $\lambda_{c}R_{c}(\xi)$, respectively, are shown by different kinds of hatched areas for which their uncertainties are mainly controlled by the errors of experimental data and the AMPT results. For more central collisions in Figure~\ref{fig-fit} (a) and (b), the contribution from coalescence is much larger than that from fragmentation in the high-$\xi$ range. With the decreasing of $\xi$, the contribution from coalescence drop down quickly while the contribution from fragmentation seems unchanged, until the two contributions become equivalent in the very low-$\xi$ range. However, it is different for the mid-central collisions in Figure~\ref{fig-fit} (c ), which shows two equivalent contributions in high-$\xi$ range and a dominant contribution from fragmentation in the low-$\xi$ range. For the most peripheral collisions in Figure~\ref{fig-fit} (d), it is hard to conclude anything due to the large uncertainties of two contributions. In general,  the effect of coalescence tends to be more dominant for high-$\xi$ range in more central collisions, while the contribution from fragmentation becomes more important for low-$\xi$ range in more peripheral collisions.

\begin{table}
\caption{\label{tab:para}The fitting parameters of $\lambda_{f}$ and $\lambda_{c}$ for different centrality bins in Pb+Pb collisions at $\sqrt{s_{_{\rm NN}}}$ = 2.76 TeV.}
\begin{ruledtabular}
\begin{tabular}{llll}
   & $\lambda_{f}$ & $\lambda_{c}$\\
    0-10\% & 0.377 $\pm$ 0.147 & 0.612 $\pm$ 0.120\\
    10-30\% & 0.346 $\pm$ 0.156 & 0.616 $\pm$ 0.131\\
    30-50\% & 0.599 $\pm$ 0.168 & 0.386 $\pm$ 0.137\\
    50-100\% & 0.379 $\pm$ 0.370 & 0.527 $\pm$ 0.338\\    
\end{tabular}
\end{ruledtabular}
\end{table}

\begin{figure}
%\hspace{-1.0cm}
\includegraphics[scale=0.45]{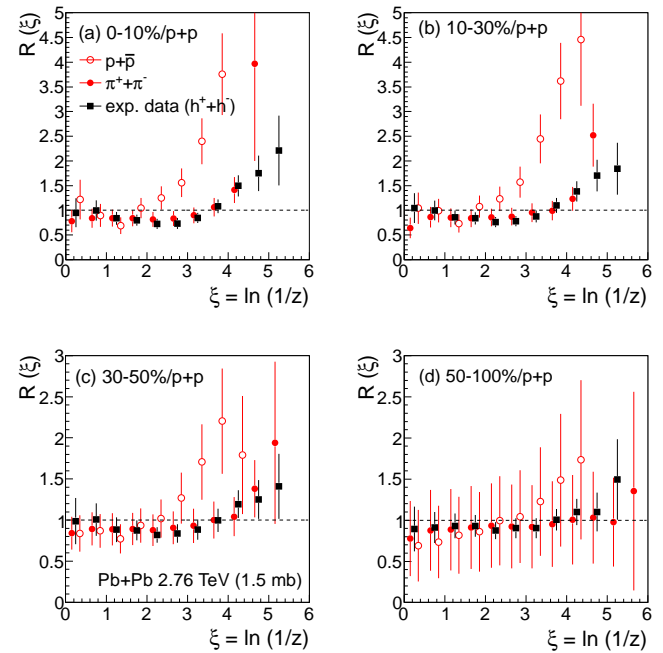}
\caption{(Color online) The jet fragmentation function ratios for charged pions (solid circles) and protons (open circles) in different centrality bins in Pb+Pb 2.76-TeV collisions to p+p collisions, where solid squares represent the data for charged hadrons from the CMS experiment~\cite{CMS:2012wxa}. Some points are slightly shifted along the $x$ axis for better representation.}
 \label{fig-predict}
\end{figure}

One basic feature of coalescence is to enhance baryon-to-meson ratio, such as $p/\pi$, because it can give a more efficient way for producing baryons than mesons. Thus it is interesting to compare the jet fragmentation function ratio for charged pions with that for protons to see the coalescence effect on jet fragmentation function. To estimate the ratios of R($\xi$) for charged pions and protons, it is simply assumed that their contribution factors, $\lambda_{f}$ and $\lambda_{c}$, are same as those for charge hadrons. The jet fragmentation functions from fragmentation and coalescence for charged pions and protons can be simulated by the AMPT model, thus the jet fragmentation function ratios for charged pions and protons can be finally estimated with Equation~(\ref{eq1}). Figures~\ref{fig-predict} (a)-(d) give the predictions about the jet fragmentation function ratios for charged pions and protons in different centrality bins in Pb+Pb collisions to p+p collisions. The ratios of R($\xi$) for charged pions are very similar to the data for charged hadrons, since most of charged hadrons are charged pions. It is interesting that the ratios of R($\xi$) for protons are significantly higher than those for charged pions especially in more central collisions though the errors are still large (which are inherited from the large uncertainties of $\lambda_{f}$ and $\lambda_{c}$). However, it is also possible that the contribution factor from coalescence $\lambda_{c}$ for protons is even larger the assumed one obtained from charged hadrons, therefore these estimated ratios of R($\xi$) for protons in Figure~\ref{fig-predict} are expected to give the lower limits to protons in the high-$\xi$ range. In addition, hard protons, dominantly produced by gluon jets~\cite{Albino:2005me}, are expected to be more suppressed than pions in the picture of jet radiative energy loss, which should bring additional suppression to the ratios of R($\xi$) for protons in the low- or intermediate- $\xi$ range. On the basis of the AMPT results, the ratios of R($\xi$) for different types of hadrons (mesons and baryons) are proposed as a good probe to study the competition between fragmentation and coalescence in jet fragmentation function. 

In summary, the jet fragmentation function is investigated based on the AMPT model with string-melting scenario.  The evolution of the jet fragmentation function suggests that it is modified not only by the strong interactions between jet parton shower and partonic medium, but also by the method of jet hadronization. However, it is hardly affected by the final hadronic rescatterings. Since different hadronization mechanisms dominate for different $p_{T}$ ranges in high-energy heavy-ion collisions, the final jet fragmentation function ratio can be decomposed into two contribution parts from fragmentation and coalescence. The results demonstrate that fragmentation competes with coalescence for jet hadronization, varying with $\xi$ and centrality. It is proposed that the comparison of the jet fragmentation function ratio between baryons and mesons is a good probe to see the effect of jet hadronization in high-energy heavy-ion collisions.

The author is grateful to R. C. Hwa for his encouragement and helpful discussions. This work was supported by the NSFC of China under Projects No. 11175232, No. 11035009, No. 11105207, No. U1232206, and No. 11220101005; the Knowledge Innovation Program of CAS under Grant No. KJCX2-EW-N01; the Youth Innovation Promotion Association of CAS; and the project sponsored by SRF for ROCS, SEM, and CCNU-QLPL Innovation Fund (Grant No. QLPL2011P01).

\end{document}